\documentclass[showpacs,amsmath,amssymb,aps,pre,11pt]{revtex4-1}
\usepackage[pdfusetitle,colorlinks=true, linkcolor=black, citecolor=black, filecolor=black, urlcolor=black]{hyperref}
\usepackage{bbold}
\usepackage{graphicx} 
\usepackage{esvect} 
\usepackage{siunitx} 
\usepackage[section]{placeins} 
\usepackage{cleveref} 
\usepackage{color}
\usepackage{soul}

\begin{document}

\title{
Geometry and design of origami bellows with tunable response 
}

\author{Austin Reid}
\email{arreid3@ncsu.edu}
\affiliation{Department of Physics, North Carolina State University, NC 27695, USA}
\author{Frederic Lechenault}
\email{flechenault@lps.ens.fr}
\affiliation{Laboratoire de Physique Statistique, Ecole Normale Sup\'erieure, PSL Research University, CNRS, 24 rue Lhomond 75005 Paris,  France}
\author{Sergio Rica}
\email{sergio.rica@uai.cl}
\affiliation{Facultad de Ingenier\'ia y Ciencias and UAI Physics Center, Universidad Adolfo Ib\'a\~nez, Avda. Diagonal las Torres 2640, Pe\~nalol\'en, Santiago, Chile.}
\author{Mokhtar Adda-Bedia}
\email{adda@lps.ens.fr}
\affiliation{Univ. Lyon, ENS de Lyon, Univ. Claude Bernard, CNRS, Laboratoire de Physique, F-69342 Lyon, France}

\date{\today}


\begin{abstract}
Origami folded cylinders (origami bellows) have found increasingly sophisticated applications in space flight and medicine. In spite of this interest, a general understanding of the mechanics of an origami folded cylinder has been elusive. With a newly developed set of geometrical tools, we have found an analytic solution for all possible cylindrical rigid-face states of both Miura-ori and triangular tessellations. Although an idealized bellows in both of these families may have two allowed rigid-face configurations over a well-defined region, the corresponding physical device, limited by nonzero material thickness and forced to balance hinge and plate-bending energy, often cannot stably maintain a stowed configuration. We have identified the parameters that control this emergent bistability, and have demonstrated the ability to design and fabricate bellows with tunable deployability.
\end{abstract}

\maketitle

\section{Introduction}
\label{sec:intro}
Origami typically calls to mind images of flowers, leaves, birds and ever more sophisticated and beautiful sculptures. Reluctant to cut, glue, or stretch their medium, artists have developed a stunning family of fold patterns and techniques to reshape flat sheets into imagined forms. Literally ``folding paper'', the art of origami is fundamentally the study of the generation of dramatic changes of a material's appearance and bulk mechanical properties via the application of a sequence of highly localized deformations.

Origami lattices are a prototypical metamaterial, readily converting an unwieldy film into a robust device capable of reliable and simple actuation~\cite{lebee2015folds}. A remarkably small fraction of a sheet is deformed when it is creased, but the mere existence of a crease dramatically changes its deformation modes. Several creases acting in concert can govern a device's  kinematics---the final device's degrees of freedom depend more on its geometry than on its local structural properties~\cite{schenk2013geometry}. Like other metamaterials, origami lattices have several exotic mechanical properties that can be tuned with small variations in their design~\cite{silverberg2014using}. Of particular interest, the Miura-ori chevron tessellation expands in all directions when pulled apart, exhibiting a negative Poisson's ratio~\cite{Wei:2013kn}. Furthermore, several classes of flat-foldable origami are known to exhibit bistable behavior~\cite{Silverberg:2015gb,Brunck:2016jr}. Origami lattices are amenable to rigorous mathematical handling~\cite{Demaine:2001uq}, which generates direction for designers exploring the space of accessible patterns. Subsequent work in that quantitative endeavor has led to a set of powerful theorems and tools, in particular to deal with the non self-intersection constraint~\cite{Belcastro:2002ku}. Classical ``rigid face'' origami tessellations are constructed with planar faces linked by flexible but well-defined hinge-like creases. These origami metamaterials have raised interest in disparate fields~\cite{Dureisseix:2012tk}, where further theoretical development is driven by a multitude of creative applications.

It is important to distinguish between origami geometry and origami mechanics. Origami geometry concerns itself with mathematically ideal objects. These bodies are inextensible, uniformly flat and are generally defined with a degree of internal symmetry. With surfaces and linkages so constrained, the base configuration of the device may be perfectly specified. Considering each crease as a hinge mechanism, it is possible to count up the degrees of freedom, which is sufficient to determine if the device is rigid or if it can be smoothly actuated. For example, the crystalline silicon cells found in deployable solar panels are effectively inelastic, and the mechanism can be analyzed much like its idealized mathematical analogue. These techniques are useful for generating starting geometries and framing problems of interest, but they are insufficient to yield mechanical insight. They are forced, by their base assumptions, to ignore the differing energetic costs of various fold configurations, and are thereby unable to predict or explain the complete mechanical response of an origami folded device. In particular, they fail for structures folded out of materials with nonzero elasticity, as both the faces and the creases have innate elastic energy. Indeed, creases have a preferred angle of repose~\cite{Lechenault:2014wn}, which can either stabilize a particular configuration or drive it far from its rigid-face equilibrium. Moreover, creases and faces tend to bend on different energy scales, and the competition of these effects leads to dramatically different behavior than what geometric models might predict. For example, Ref.~\cite{Silverberg:2015gb} demonstrates how face bending can generate a pathway for an origami mechanism to follow while transitioning through a geometrically forbidden configuration to a lower energy state, while in Ref.~\cite{Lechenault:2014wn} a straightforward technique to measure the competition between crease and plate bending is demonstrated.

Much of previous theoretical work was focused on planar lattices~\cite{Wei:2013kn}, although cylindrical configurations are of considerable technical interest. Origami folded cylinders have found applications in space technology for deployable sails and booms~\cite{Schenk:2014jc,Guest:1994uo}, medical devices such as stents~\cite{Kuribayashi:2006wo}, and even nuclear physics~\cite{Austin}. Although these fields are very sensitive to reliability and cost concerns, pattern development has been conducted in a largely ad-hoc manner~\cite{Schenk:2013to} due to the absence of a general predictive framework for their performance. Recent work with such configurations has led to several remarkable theoretical developments. In Ref.~\cite{Tachi:2012wl}, a family of rigid foldable cylindrical bellows is identified and a mechanism whereby the mechanics of such bellows could be tuned is demonstrated~\cite{Yasuda:2015vy}. It was later proved that a cylinder constructed of radially arranged Miura corrugations is incapable of rigid foldability~\cite{Bos:2015uu}. Recent developments have explored the bistability of bellows patterned with Miura-ori folds~\cite{Cai:2015hb} as well as a Kresling pattern~\cite{Jianguo:2015ic}. These works utilize an elastic rod framework to explore the dynamic response of a folded bellows. This suffices to illustrate the existence of geometrically allowed bistable configurations, but it fails to capture the behavior of a folded device as it actuates.

In the present work, we use properties established by a general solution for allowable configurations to predict and explain responses of real bellows. To this purpose, we will first define and solve the geometry at hand, detailing each constraint and assumption. We will then explore the mechanical response of their physical manifestations, necessitating the construction of a series of tunable cylindrically symmetrical bellows before subjecting them to controlled actuation and collapse. We will finally conclude with a discussion of the behavior observed during actuation, with a special focus on its possible applications.

\section{Origami bellows Geometry}
\label{sec:geometry}

Given an ideal origami pattern represented as a system of linked rigid polyhedra, there exist multiple families of frameworks that may be used to analyze its allowable configurations. If an origami system is represented as a mechanism of rigid linkages~\cite{Guest:1994uo}, it can be subjected to classical constraint-counting techniques. Some of these frameworks invoke quaternionic algebra to generate the relative rotations of their linkages~\cite{Guest:1994uo} or faces~\cite{Wu:2010gq}. Other treatments of Miura-ori sheets and cylinders use the angle between faces as their control parameters~\cite{Wei:2013kn,Wu:2010gq}. In the following, we instead parameterize origami tessellations using the set of vectors associated with the crease network~\cite{Brunck:2016jr}. By solving the fully-constrained behavior of a periodic fundamental origami cell, we have found an analytic solution for all possible rigid-face states accessible from both cylindrical Miura-ori and Kresling patterns.

\begin{figure}[!htbp]
\includegraphics{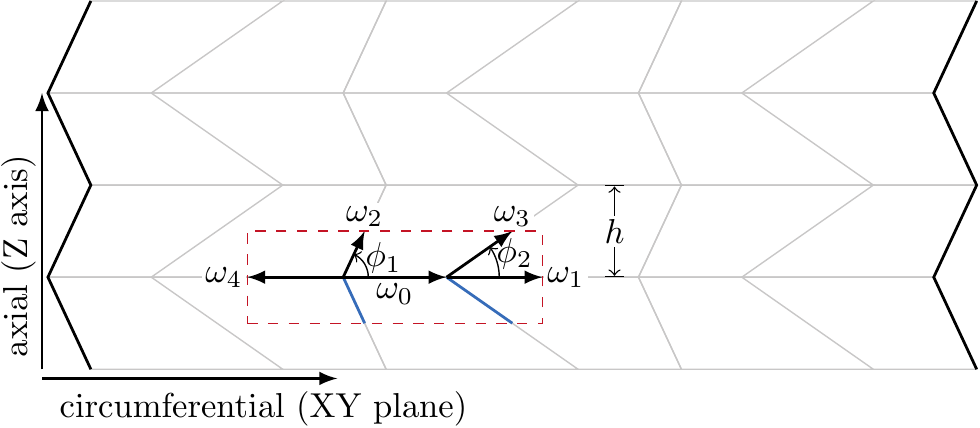}
\caption{Cylindrical Miura Ori corrugation: thick left and right edges connect to one another. The framework used in this paper does not specify whether edges are valley or mountain folded, a property that instead emerges from the solution of its geometry. The red dotted rectangle indicates a unit cell. The lower blue edges are implicitly assumed to mirror $\omega_2$ and $\omega_3$ about the plane established by $\omega_0$ and $\omega_1$. A single band of circumferential rigid panels is defined by its lateral size $h$.}
\label{fig:miura:fullgrid:mother}
\end{figure}

\begin{figure}[ht]
\includegraphics{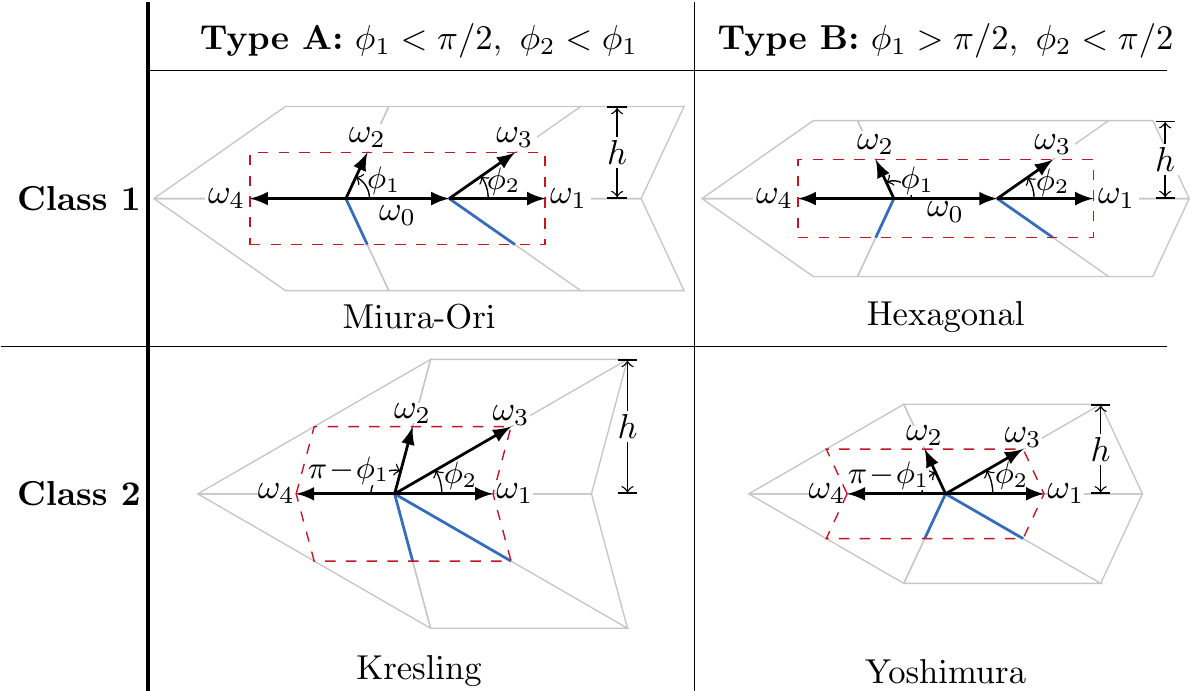}
\caption{Various unit cells (delimited by red dashed rectangles) derived from the Miura-ori folding patterns. A global scaling factor determines the length of the unit cell. At least three circumferential cells are required for making a bellows.}
\label{tab:mother:alltypes}
\end{figure}

To illustrate the benefits of this method, consider the folding pattern depicted in~\cref{fig:miura:fullgrid:mother}. As shown in table~\ref{tab:mother:alltypes}, the tessellations of several types of rotationally symmetric bellows can be derived from variations on this unit cell. Planar tilings of these unit cells have been studied extensively, but relatively few works have studied cylindrical configurations \cite{Jianguo:2015ic,Bos:2015uu}. Some developable patterns on the cylinder lack rotational symmetry~\cite{Yasuda:2015vy}, but here we will focus on regular cylindrical tilings of the above aforementioned cells, whose regularity will be formally introduced as uniform rotational symmetry.

\begin{figure}[!htbp]
\includegraphics[height=2.in]{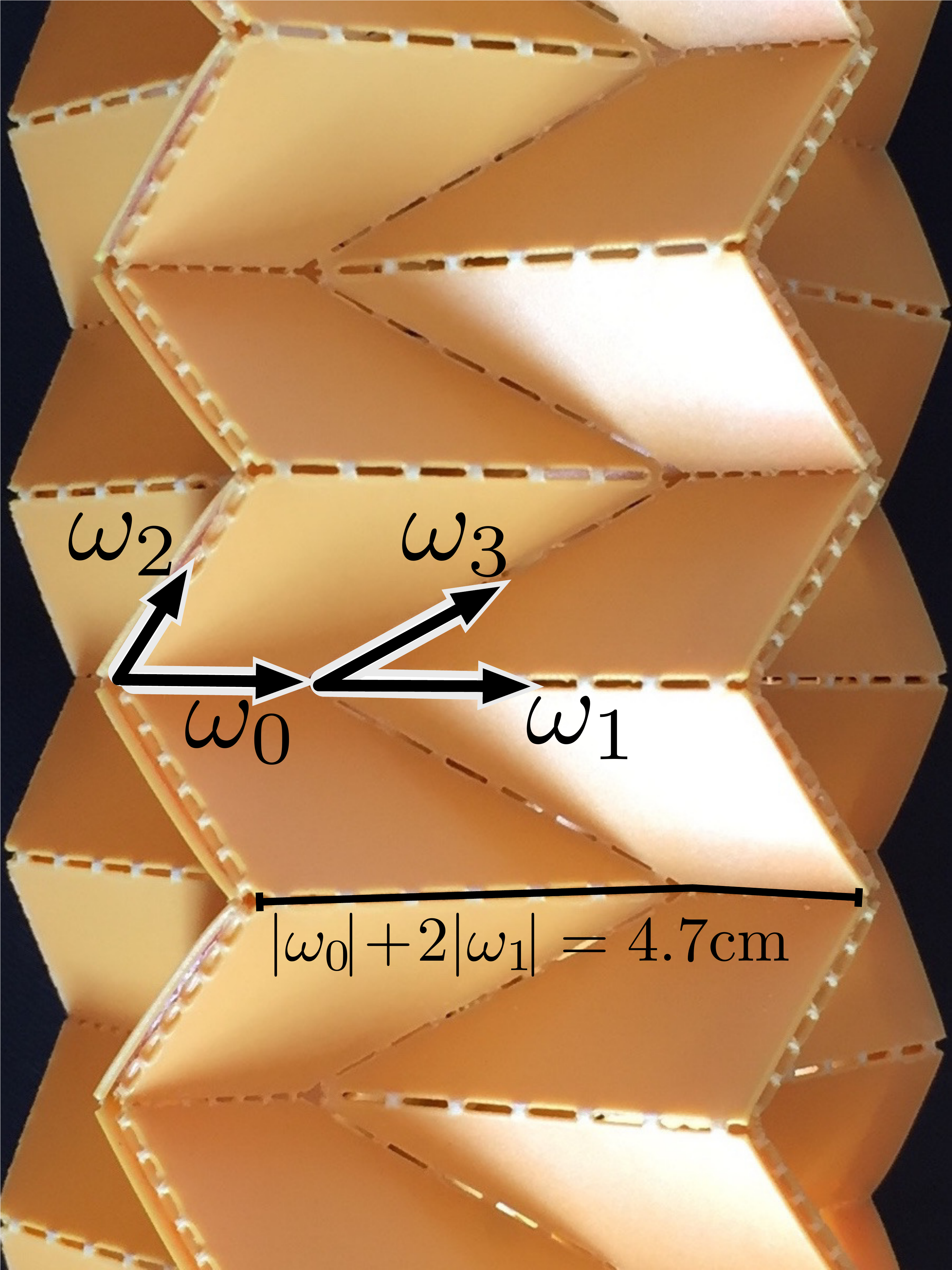}
\includegraphics[height=2.in]{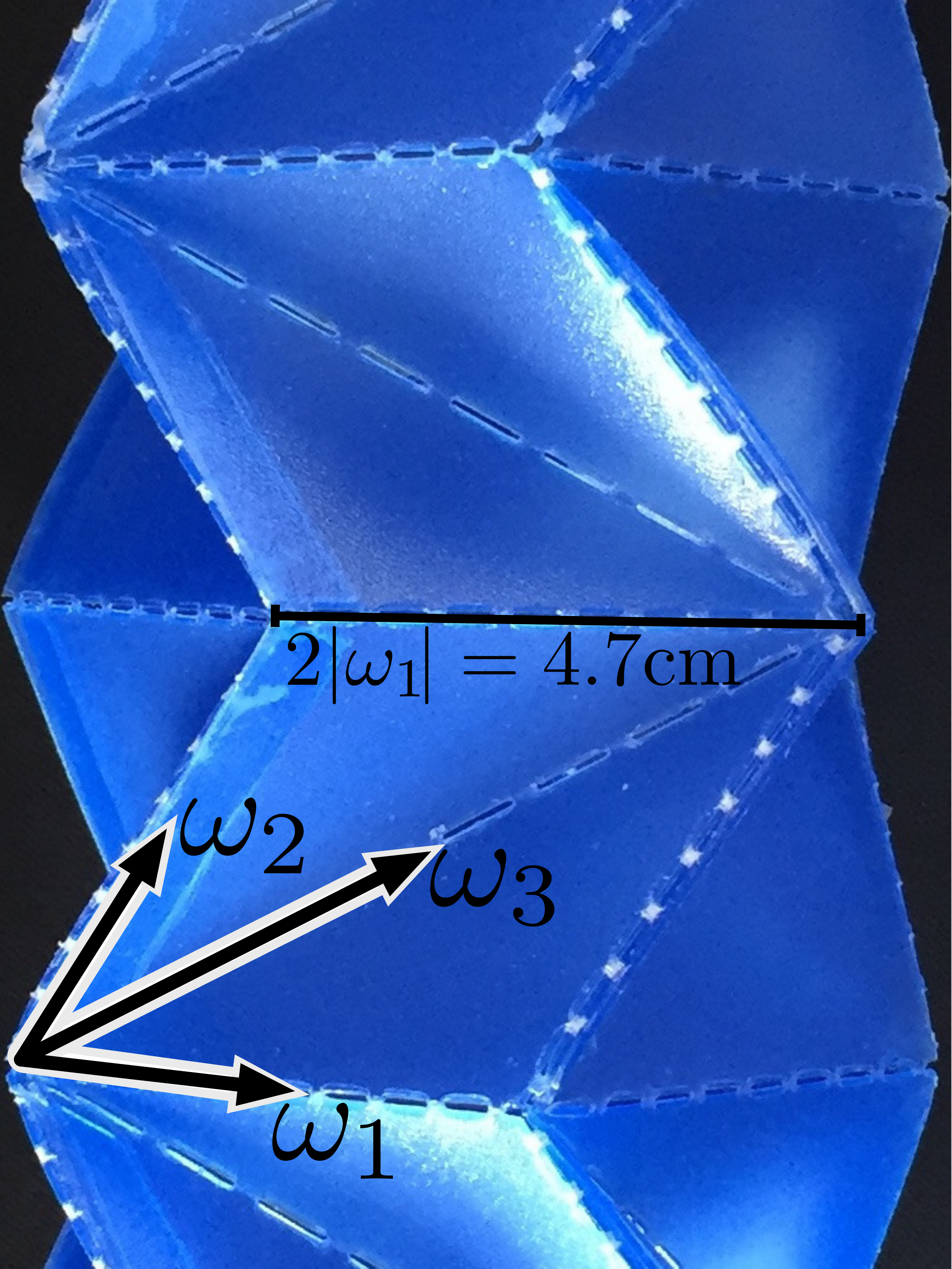}\\
\includegraphics[width=1.5 in]{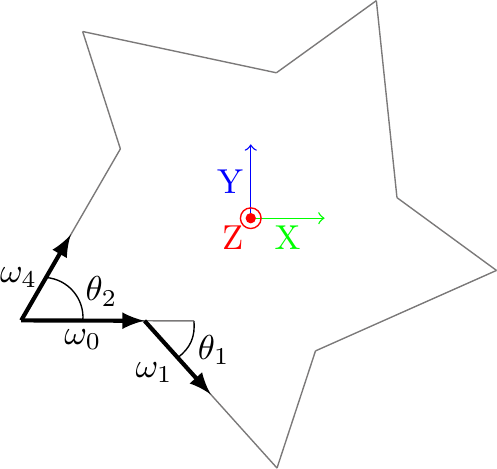}
\includegraphics[width=1.5 in]{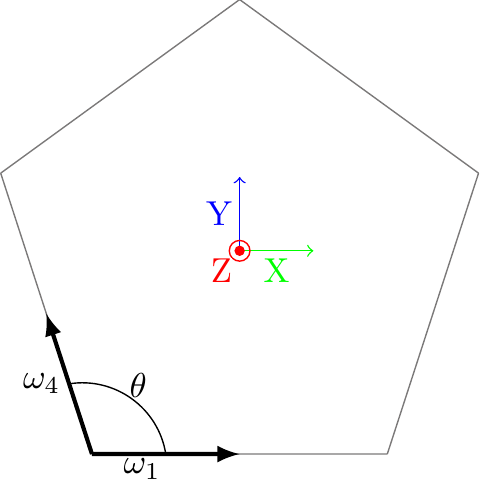}
\caption{First row: Miura-Ori (left) and Kresling (right) bellows models annotated with example unit cell vectors. Second row: annotated vector illustration of unit cell in XY plane. Photographs are at approximately at the same scale and these structures correspond to $n=5$ unit cells.}
\label{fig:miura:kresling:pictures:labeled}
\end{figure}

The unit cells illustrated in~\cref{tab:mother:alltypes} generate bellows like those shown in \cref{fig:miura:kresling:pictures:labeled}. They have been organized by class and type leading to four major domains. Type~A and type~B differ in the value of the continuously variable angle \(\phi_1\), where type~A patterns correspond to \(\phi_1<\frac{\pi}{2}\) and type~B patterns are characterized by \(\phi_1>\frac{\pi}{2}\). Class~1 and 2 patterns differ in the existence of the crease vector \(\vv{\omega}_0\) which leads to a different set of fold parameterizations and constraint counts, but will not dramatically alter the solutions for rigid-face states. When \(0<\phi_2<\phi_1<\frac{\pi}{2}\), the generated pattern is the familiar Miura corrugation for class~1 and Kresling pattern for class~2. If \(\frac{\pi}{2}<\phi_1<\pi\) and \(0<\phi_2<\frac{\pi}{2}\), a hexagonal grid is generated instead, for class~1 and Yoshimura pattern for class~2. These angular ranges are nearly sufficient to define the class of figures, but one additionally needs to maintain positive edge lengths. As far as class~2 fold patterns are concerned, thin-walled cylinders under axial compression tend to develop a rigid triangular tessellation known as a Yoshimura pattern, first described in~\cite{Yoshimura:1955up}. This tessellation is described by this framework as class~2--type~B. Under torsion, thin-walled cylinders of certain lengths buckle into a twisted triangular configuration, known as a Kresling pattern~\cite{Hull:2002wn}, described here as class~2--type~A.

\Cref{fig:miura:fullgrid:mother} shows that a class~1 unit cell is described by a total of five fold vectors \(\vv{\omega}_i\), with \(|\vv{\omega}_i|=\ell_i\). By setting \(\ell_0+\ell_1+\ell_4=1\), a global length scale is defined allowing the parameterization of all edge lengths as:
\begin{align}
\left|\vv{\omega}_0\right| &= \ell_0\;,&
\left|\vv{\omega}_1\right| = \left|\vv{\omega}_4\right| &= \ell_1 = \frac{1}{2}\left(1-\ell_0\right) \;,\\
\left|\vv{\omega}_2\right| &= \ell_2 = \frac{h}{2} \csc\phi_1 \;,&
\left|\vv{\omega}_3\right| &= \ell_3 = \frac{h}{2} \csc\phi_2\;.
\end{align}
The free length parameters \(\ell_0\) and and lateral size of a single band \(h\) are geometrically restricted to the following domains:
\begin{align}
0<h&\le \frac{1-\ell_0}{\cot\phi_2-\cot\phi_1}\;.\label{eq:h:domain}
\end{align}
Inequality~(\ref{eq:h:domain}) ensures that \(\ell_1\) remains positive and that the edge defined by \(\vv{\omega}_3\) does not cross  \(\vv{\omega}_2\)  in an adjacent unit cell. Note that class~2 fold patterns are described by this same framework but with \({\ell_0=0}\) and \(h = 1/(\cot\phi_2-\cot\phi_1)\).

\begin{figure}[!htbp]
\includegraphics{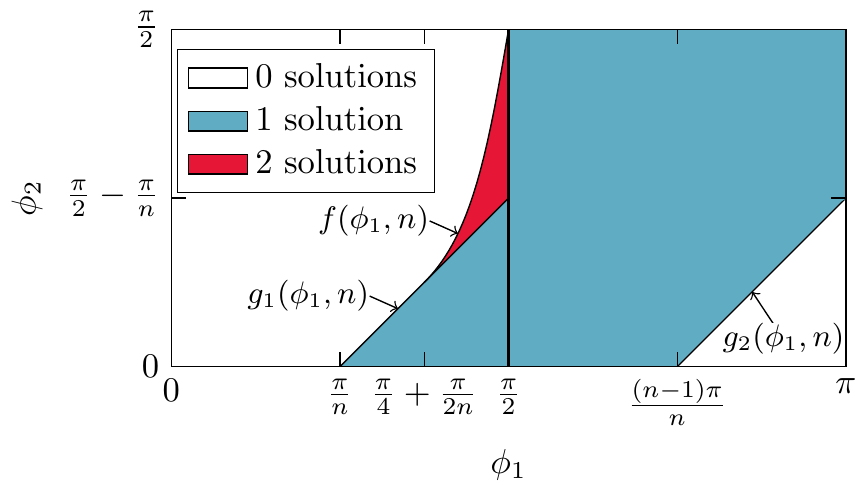}
\caption{Parameter space for class~1 and class~2 family bellows for a number \(n\geq3\) of unit cells.}
\label{fig:phase-diagram}
\end{figure}

When the imprinted network of creases is rigidly folded, two degrees of freedom can be fixed because of global rotational symmetry. A first degree of freedom can be eliminated by taking \(\vv{\omega}_0=\ell_0 \hat{\imath}\), where \(\hat{\imath}\) is a unit vector parallel to the X-axis. Second, the vectors \(\vv{\omega}_1\) and \(\vv{\omega}_4\) also lie in the XY plane, locking the final degree of freedom and expediting the generation of the remaining vectors. Here, we utilize \(SO(3)\) rotation matrices in lieu of \citeauthor{Wu:2010gq}'s quaternionic approach~\cite{Wu:2010gq}.
\begin{align}
\vv{\omega}_1 &= \ell_1 R_z(-\theta_1)\,\hat{\imath}\;,\label{eq:genw1}\\
\vv{\omega}_2 &= \ell_2 R_x(\psi)\,R_y(-\phi_1)\,\hat{\imath}\;,\label{eq:genw2}\\
\vv{\omega}_3 &= \ell_3 R_x(\psi)\,R_y(-\phi_2)\,\hat{\imath}\;,\label{eq:genw3}\\
\vv{\omega}_4 &= \ell_4 R_z(\theta_2)\,\hat{\imath}\;,\label{eq:genw4}
\end{align}
where \(\theta_1\) describes the angular deflection of \(\vv{\omega}_1\) with respect to \(\vv{\omega}_0\) and \(\theta_2\) parameterizes the opening angle between \(\vv{\omega}_0\) and \(\vv{\omega}_4\), as seen in \cref{fig:miura:kresling:pictures:labeled}. To handle the opening angle of pattern about \(\omega_0\), the angular deflection \(\psi\) from vertical ($z$-axis) is a measure of the panel's angular deviation (\({\pi-2\psi}\) being the opening angle between adjacent bands of folds). To ensure a symmetric cylindrical configuration of class~1 constructions, we introduce an additional constraint on \(\theta_1\) and \(\theta_2\).
\begin{equation}
\theta_2 = \pi-\frac{2\pi}{n}-\theta_1\;.
\label{eq:thetas}
\end{equation}
Here $n$ is the number of unit cells. Condition~(\ref{eq:thetas}) is necessary to generate well-behaved closed tubes for all \(n\ge3\). It is easily shown that the parameterization described by \crefrange{eq:genw1}{eq:genw4} reduces to a system of two equations, with the as-yet unspecified parameters \(\theta_1\) and \(\psi\):
\begin{align}
\vv{\omega}_1\cdot \vv{\omega}_3 = \ell_1\ell_3\left[\cos\theta_1\cos\phi_2+\sin\theta_1\sin\phi_2\sin\psi\right] &= \ell_1\ell_3 \cos\phi_2 \;,\label{eq:miuraconstraint:w13}\\
\vv{\omega}_2\cdot \vv{\omega}_4 = \ell_2\ell_4\left[\cos\theta_2\cos\phi_1-\sin\theta_2\sin\phi_1\sin\psi \right]&= -\ell_2\ell_4\cos\phi_1 \;.\label{eq:miuraconstraint:w24}
\end{align}
Using \cref{eq:thetas}, one can show that this system of equations is satisfied by
\begin{align}
\tan{\frac{\theta_1}{2}}& = \frac{1}{2\tan\frac{\pi}{n}}\left[1-\frac{\tan\phi_2}{\tan\phi_1} \pm \sqrt{
  \left(\frac{\tan\phi_2}{\tan\phi_1}-1\right)^2-4\frac{\tan\phi_2}{\tan\phi_1}\tan^2\frac{\pi}{n}
}
\right]\;,
\label{eq:solution:tantheta}\\
\sin\psi &= \frac{\tan\frac{\theta_1}{2}}{\tan\phi_2}\;.
\label{eq:rawsimpletheta1}
\end{align}
Equations~(\ref{eq:solution:tantheta}) and (\ref{eq:rawsimpletheta1}) determine $\theta_1$ and $\psi$ as functions of $\phi_1$ and $\phi_2$, the only control parameters of bellows geometry. One can show that the solution of~\cref{eq:solution:tantheta} satisfies \(\theta_1\le\pi-\frac{2\pi}{n}\), and thus self-intersection of the faces is implicitly avoided. However for \cref{eq:solution:tantheta} and \cref{eq:rawsimpletheta1} to yield physically meaningful solutions, one must still satisfy that \(\theta_1\) is real and \(|\sin\psi|\le1\). Using these conditions one can show that there are no more than two geometrically allowed rigid face configurations for a closed band constructed of at least 3 unit cells. The structure of the phase diagram is illustrated in \cref{fig:phase-diagram}. The regions of the \(\phi_1,\phi_2\) parameter space where physical solutions exist are bounded by several functions:
\begin{align}
f(\phi_1,n) &= \arctan\left(\tan(\phi_1)\cdot\frac{1-\sin\frac{\pi}{n}}{1+\sin\frac{\pi}{n}}\right)\;,\label{eq:f:phi1}\\
g_1(\phi_1,n) &= \phi_1-\frac{\pi}{n}\;,\label{eq:g1}\\
g_2(\phi_1,n) &= \phi_1-\pi+\frac{\pi}{n}\;.\label{eq:g2}
\end{align}
\(f(\phi_1,n)\) is found by limiting the solutions of \cref{eq:solution:tantheta} to be real and \(g_1\) (resp. \(g_2)\) corresponds to the case \(\sin\psi=1\) (resp. \(\sin\psi=-1\)).

Although the number of faces is identical, class~1 and class~2 tessellations differ in their total number of folds requiring careful verification of allowable rigid-face configurations for class~2 folded patterns. Using the same approach one can show that the corresponding solution space is indeed also bounded by \cref{eq:f:phi1,eq:g1,eq:g2}, generating identical diagrams as the solutions detailed in \cref{fig:phase-diagram}, indicating that the bistability of these bellows is topological in nature. These results dovetail with Connelly's Bellows Theorem~\cite{Connelly:1997tw}, which states that continuous deformations of a closed triangulated surface cannot change its volume, and that the number of enclosed volumes attainable by reconfiguring a closed triangulated surface is finite. By assuming rotational symmetry, this solution does not explore the ability for a single band of folds to continuously deform. These folding modes require $\omega_{0}$, $\omega_{1}$, and $\omega_{4}$ to not be coplanar, a folding mode which is blocked by the presence of the cells corresponding to $\omega_{2}$ and $\omega_{3}$'s mirrored edges.

The classification by type in \cref{tab:mother:alltypes} is chosen to mirror the structure of the parameter space in \cref{fig:phase-diagram}. As there are zero free parameters in the system of equations given by \cref{eq:solution:tantheta,eq:rawsimpletheta1}, we will only find a finite number of allowable solutions at any point in parameter space. The figures constructed with \(\phi_1>\frac{\pi}{2}\) are at best monostable. The boundary defined by \(g_2(\phi_1,n)\) generates a flat-folded monostable figure.

Type~A bellows are the one generated on the left half of \cref{fig:phase-diagram}, with a bistable region neighboring a larger monostable region. The bistable region is bounded on the left by \(f(\phi_1,n)\) and below by \(g_1(\phi_1,n)\). Along the curve \(g_1\) one of the configurations is flat-folded. At \((\phi_1,\phi_2)=(\frac{\pi}{2},\frac{\pi}{2}-\frac{\pi}{n})\) the deployed state is completely extended, with perfectly flat walls along the bellows' axial direction. Moving to smaller values of \(\phi_1\) along \(g_1\), the deployed configuration's hinge angle decreases until \({\phi_1=\frac{\pi}{4}+\frac{\pi}{2n}}\), at which point the bistable states are degenerate. The deployable bellows explored in \cite{Schenk:2013to} all lie along \(g_1\), with increasing deployability as \(\phi_1\) increases from \(\frac{\pi}{4}+\frac{\pi}{2n}\) to \(\frac{\pi}{2}\).

\begin{figure}[!htbp]
\includegraphics{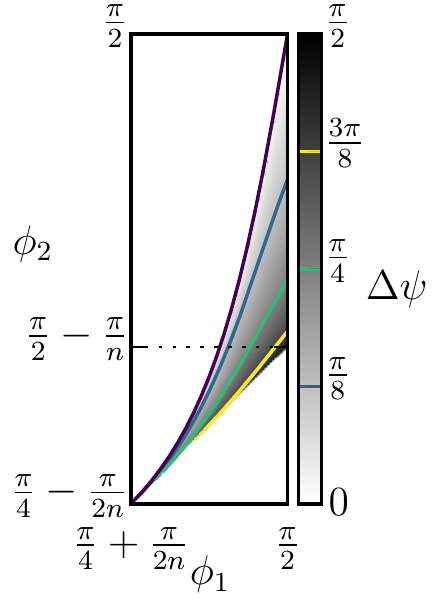}
\caption{Detailed figure of the (red) bistable region of \cref{fig:phase-diagram}. Greyscale shading indicates the angle difference \(\Delta\psi\) (in radians) between collapsed and deployed solutions, with colored call outs for the contours indicated on the figure's right. The purple curve represents zero separation between the solutions, indicating the solution's degeneracy along the curve \(f(\phi_1,n)\).}
\label{fig:deploymentpsi}
\end{figure}

In the region of existence of two states, the difference between more collapsed and more extended configurations for an origami cylinder may be colloquially referred to as its ``deployability''. The magnitude of extension between the two states is easiest to measure from the angle difference \(\Delta\psi\) between the two allowed rigid face configurations by \cref{eq:solution:tantheta} and \cref{eq:rawsimpletheta1}. \Cref{fig:deploymentpsi} illustrates the parameter space governing these tubes. Along the line \(\phi_2 = \phi_1-\frac{\pi}{n}\), one of the solutions is flat foldable and therefore lies entirely within the XY-plane, and along the line \(\phi_1=\frac{\pi}{2}\), one of the solutions is maximally extended, perpendicular to the XY-plane. The pair of states found at \({\left(\phi_1,\phi_2\right)=\left(\frac{\pi}{2},\frac{\pi}{2}-\frac{\pi}{n}\right)}\) therefore correspond to flat foldable and maximally extended states. This point is the maximally deployable configuration, with \(\Delta\psi=\frac{\pi}{2}\). The deployability decreases monotonically as one moves away from this maximal configuration. This overall behavior is independent of the number of facets in the cylinder's folding pattern.
\Cref{fig:deploymentpsi} also hints that small variations in \(\phi_1\) or \(\phi_2\) can have dramatic changes in the mechanical stability of these objects.

Individual bands of a class~2 tessellation do not affect the configuration of their neighbors because the angle parameter for the polygonal footprint, \({\theta=\pi - \frac{2 \pi}{n}}\), does not vary during the pattern's collapse (see \cref{fig:miura:kresling:pictures:labeled}). This behavior is different from that of a class~1 folded cylinder: \(\theta_1\) and \(\theta_2\) change from one stable configuration to the other, so each band of the bellows is forced to move in lockstep with its neighbors. While both the individual bands have the same allowable configurations, the deployment behavior of an aggregate bellows should not be the same. Each stage of a Kresling bellows is able to snap between its bistable states independently of its neighbors, but a Miura-ori cylinder's only stable rigid face configurations are completely collapsed or completely deployed. This lack of intermediate stable configurations contributes to the appearance of smooth deployability. To demonstrate this most clearly, a physical bellows should be constructed.

\section{Mechanics}
\subsection{Device fabrication}
\label{sec:device:fabrication}

In order to generate regularly folded bellows, a laser cutter was used to etch flat panels of a substrate, which could then be assembled into a uniform cylinder. Cylinders were initially constructed by cutting a complete fold pattern from a single sheet that included a row of tabs with which the sheet was glued into a cylinder. Once glued, the panels' hinges could be folded so that the entire device collapsed into its preferred configuration. While useful for exploring various fold patterns, this assembly technique is not acceptable for device fabrication. The height of the energy barrier between bistable configurations is a function of the face bending energy~\cite{Silverberg:2015gb}, and this technique, although rapid, generates a single column of particularly stiffer walls.

\begin{figure}[!htbp]
\includegraphics{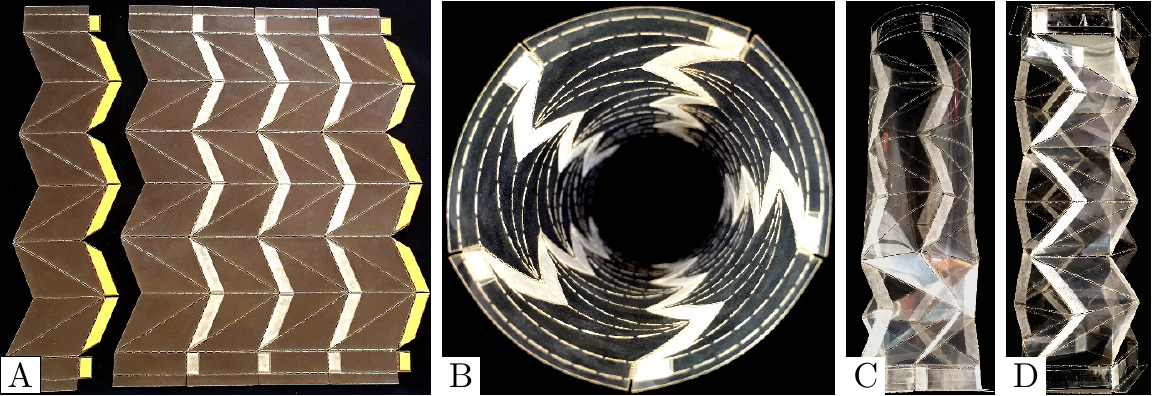}
\caption{Isotropic bellows construction. Individual strips are patterned with the laser cutter before being attached by the small tabs seen at right-most edge of (A). Strips are then rolled into a tube (B) which is then collapsed one cell at a time (C) to form the final bellows (D)}
\label{fig:paperlantern:making}
\end{figure}

Instead, our bellows are constructed much like a glued paper lantern. Individual strips of acetate sheets with thickness \SI{.1}{mm}  (transparency slides) are cut out. Instead of scoring creases, the sheets are perforated: leaving a constant length fraction attached for each crease ensures a hinge energy that scales appropriately with the length of the crease. Each strip has a set of tabs which connect it to its neighbor with double-sided tape (\cref{fig:paperlantern:making}A). Panels are aligned where laser-cut lines converge, and global alignment is double-checked with the flatness of the assembled sheet of strips before they are connected into a tube. This generates a device with an isotropic cross-section, as seen in \cref{fig:paperlantern:making}B. Once shaped into a tube, the top and bottom regions are reinforced with a layer of scotch tape. Without this tape, the tubes tend to pull apart from the ends under the stress of the initial folding. Edges are gently pre-creased, working along the entire tube (\cref{fig:paperlantern:making}C). As the tube approaches the desired shape, more force is used until the panels buckle flat (\cref{fig:paperlantern:making}D). With this design, one edge of each face is stiffer than the remainder of the bellows. Fortunately, the stiffened region is very close to the crease, where mechanics are dominated by the hinge energy. The bellows is collapsed during assembly, giving a preliminary indication of how it will respond to forced cycling.

The naming convention used in \cref{tab:mother:alltypes} may be adapted for device fabrication by considering a single band of circumferential rigid panels (see \cref{fig:miura:fullgrid:mother}). A class~2 (triangular tessellation) bellows can be constructed with any number of these bands.  An odd number of bands must twist during compression, but mirrored pairs of bands will have no net twist. A class~1 (Miura-ori) bellows should be designed with an even number of bands, as each one needs to be paired with its mirror image. In order to average out small defects from construction while keeping the bellows as large as possible, each bellows is constructed with 6 total bands per strip. In addition to the bands themselves, a bit of extra material is required to seal the end of the bellows. Furthermore, a transition pattern is required for a Miura-Ori folded bellows to connect smoothly to these convex and rigid end caps. Finally, five-panel ($n=5$) bellows were selected in the following study.

Although Miura-ori and triangular tessellated unit cells are described by identical phase diagrams, the aggregate behavior of a collection of unit cells may be different during extension and collapse. Exploring this divergent behavior requires the careful selection of a control parameter to hold constant among multiple test geometries. Because the total number of unit cells will vary during testing, the total fold length and the folds length relative to the overall base size were conserved. 

\subsection{Experimental Observations}
\label{sec:data:observations}

Because the bellows are fabricated with nonzero hinge energy and panels of finite thickness, a mathematically bistable configuration may not be mechanically bistable, i.e. able to remain in one metastable state without external forcing. As a matter of fact, \cref{fig:phase-diagram} shows that bellows constructed along the curve \(f(\phi_1,n)\) are degenerate and thus monostable, and bellows constructed along the segment (\(\phi_1=\pi/2,\frac{\pi}{2}-\frac{\pi}{n}\leq\phi_2\leq\frac{\pi}{2}\)) have a barrier to their collapse, and are thus mechanically bistable. Therefore, one expects that a transition between these regimes should exist for a critical design angle $\phi_1=\phi_c$. When \(\phi_1<\phi_c\), the bellows is unable to remain in its collapsed state and is expected to deploy itself to very near its rigid-face extended state after its collapse. This property provides with a large design space for actuation, as will be discussed below.

Inspired by~\cite{Schenk:2013to,Kane:2000tf}, initial values of \(\phi_1=\frac{2\pi}{5}\) and \(\phi_2=\phi_1-\frac{\pi}{n}\) were selected. This generates a flat-foldable bellows that is very close to the minimal distortion geometry described in~\cite{Kane:2000tf}. We found that such a ``flat foldable'' bellows is unable to maintain a stowed configuration, instead demonstrating a self-deployability. Although it would be possible to move into the non-flat stowed region (\(\phi_2 > \phi_1-\frac{\pi}{n}\)), it is more interesting to explore various values of \(\phi_1\) by keeping \(\phi_2 = \phi_1-\frac{\pi}{n}\).  The exact value of the critical \(\phi_1\) for this transition is dependent on assembly, but it appears that \(\phi_c\gtrsim \frac{2\pi}{5}=\ang{72}\). Class~1 and 2 bellows were designed on either side of this transition (\(\phi_1=\ang{68}\) and \ang{76}) in order to demonstrate its existence.

\begin{figure}[!htbp]
\includegraphics[width=0.5\textwidth]{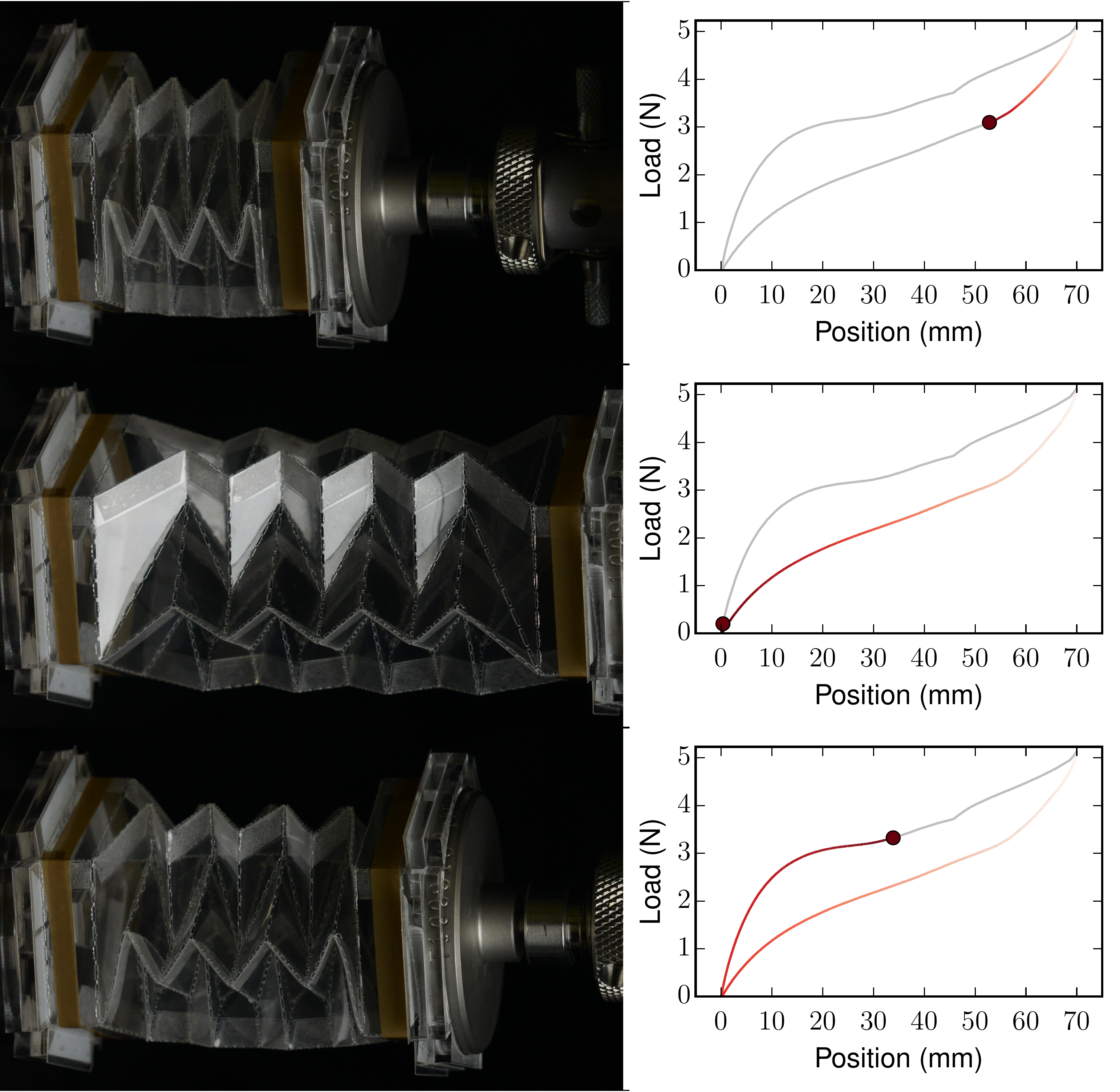}
\caption{Class~1--Type~A (Miura-ori) bellows with \(\phi_1=\ang{68}\). This device collapses smoothly, with only a small step as its transition panels collapse. Force measurement corresponding to image on left is indicated as a solid point. The colored gradient trailing the ball indicates previous force/position measurements, and the shadowed line indicates future measurements. See also Movie-S1 in~\cite{Note1}.}
\label{fig:bellows:phi68:miura}
\end{figure}

\begin{figure}[!htbp]
\includegraphics[width=0.5\textwidth]{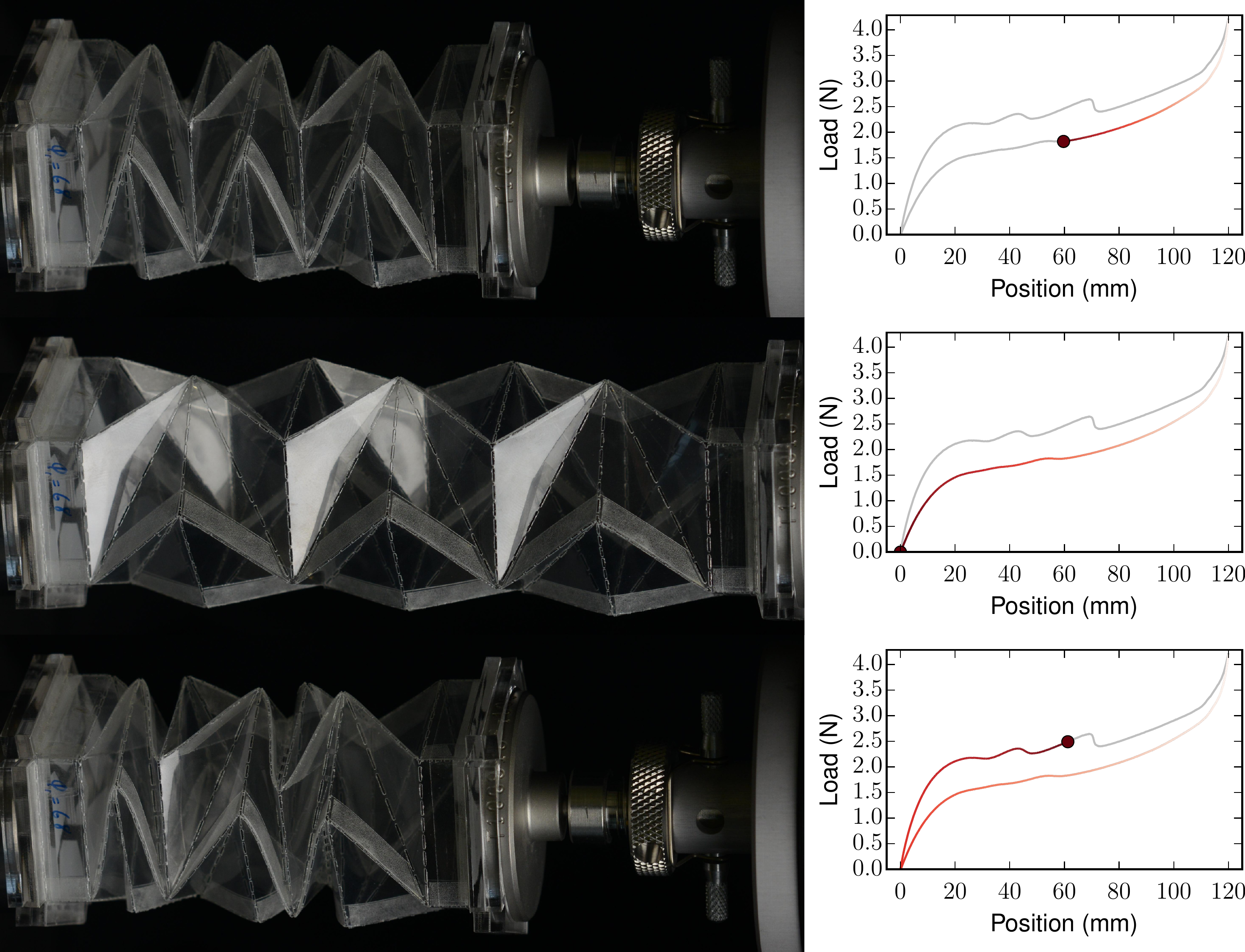}
\caption{Class~2--Type~A (Kresling) bellows with \(\phi_1=\ang{68}\). This device is constructed with identical angles \(\phi_1\) and \(\phi_2\) as \cref{fig:bellows:phi68:miura}, yet it collapses with small steps as each paired panel crumples. It deploys smoothly without external forcing. See also Movie-S2 in~\cite{Note1}.}
\label{fig:bellows:phi68:tri}
\end{figure}

\begin{figure}[!htbp]
\includegraphics[width=0.5\textwidth]{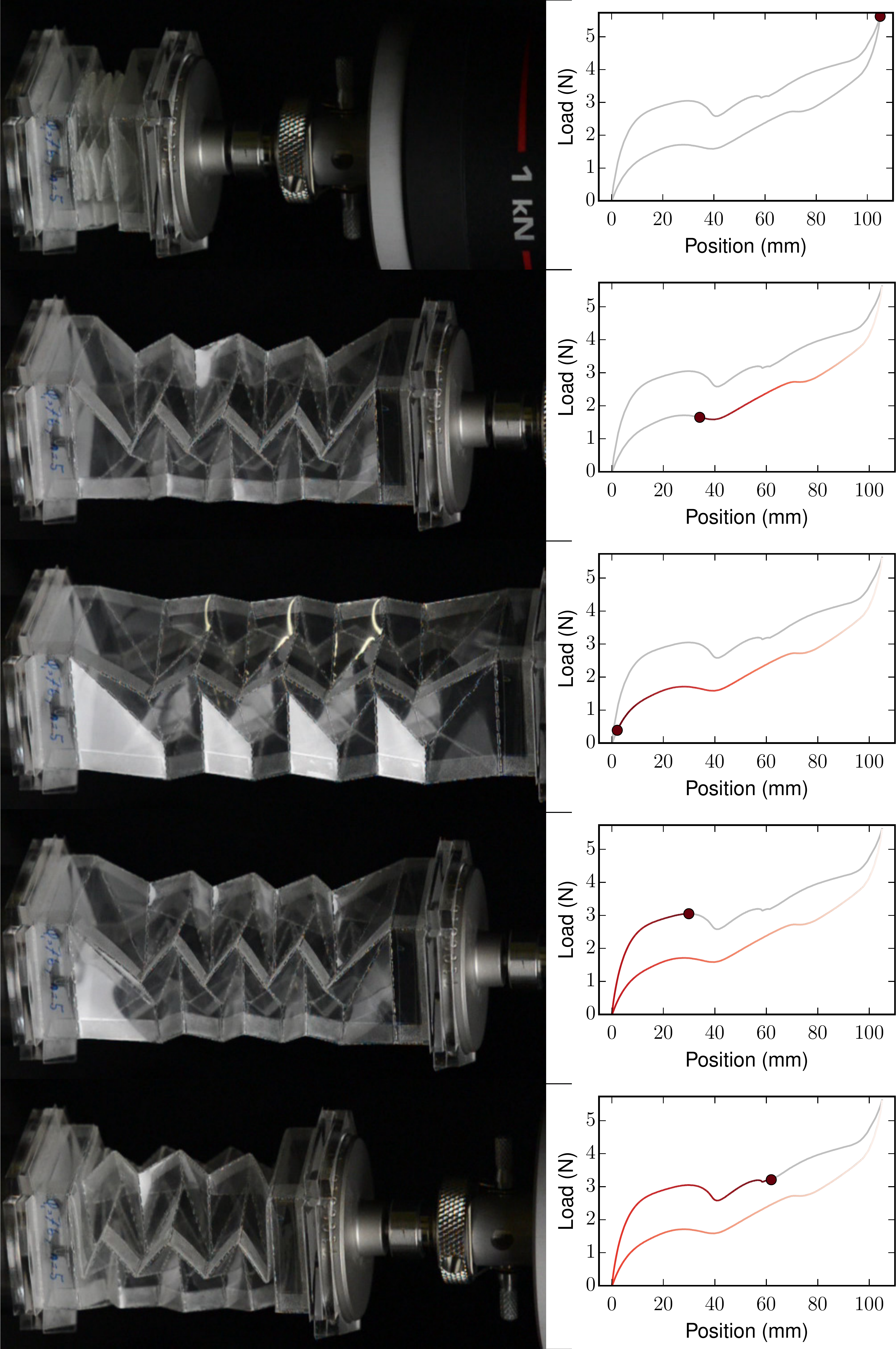}
\caption{Class~1--Type~A (Miura-ori) bellows with \(\phi_1=\ang{76}\). This device collapses with a single large step at its transition panels, and deploys in a fairly smooth manner. See also Movie-S3 in~\cite{Note1}.}
\label{fig:bellows:phi76:miura}
\end{figure}

\begin{figure}[!htbp]
\includegraphics[width=0.5\textwidth]{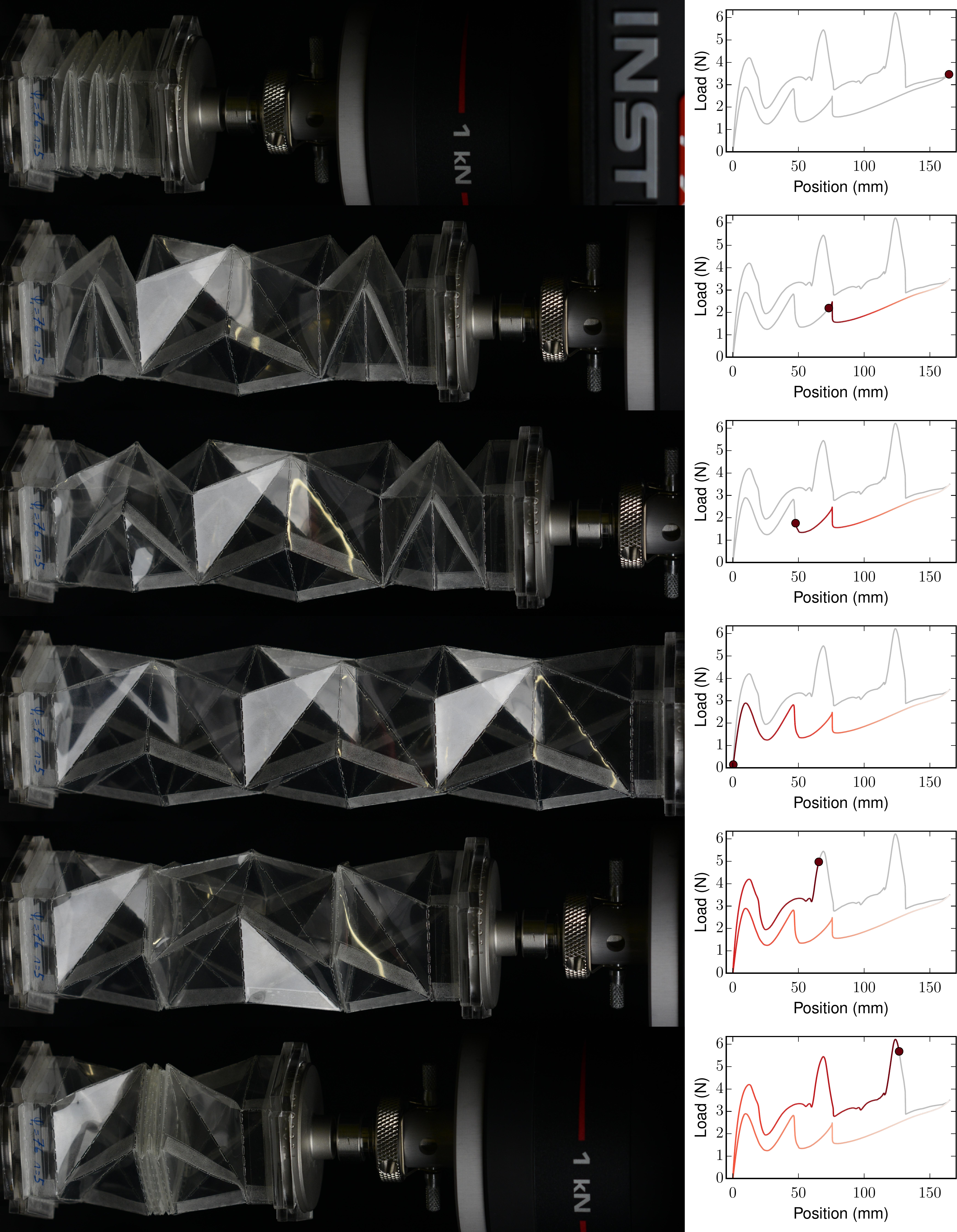}
\caption{Class~2--Type~A (Kresling) bellows with \(\phi_1=\ang{76}\) shows the most dramatic behavior of the lot. Contrary to class~1--type~A shown in \cref{fig:bellows:phi76:miura}, this bellows must be pulled apart at each step, and snaps shut as it collapses with increasing rapidity. See also Movie-S4 in~\cite{Note1}.}
\label{fig:bellows:phi76:tri}
\end{figure}

\begin{figure}[!htbp]
\includegraphics[width=0.5\textwidth]{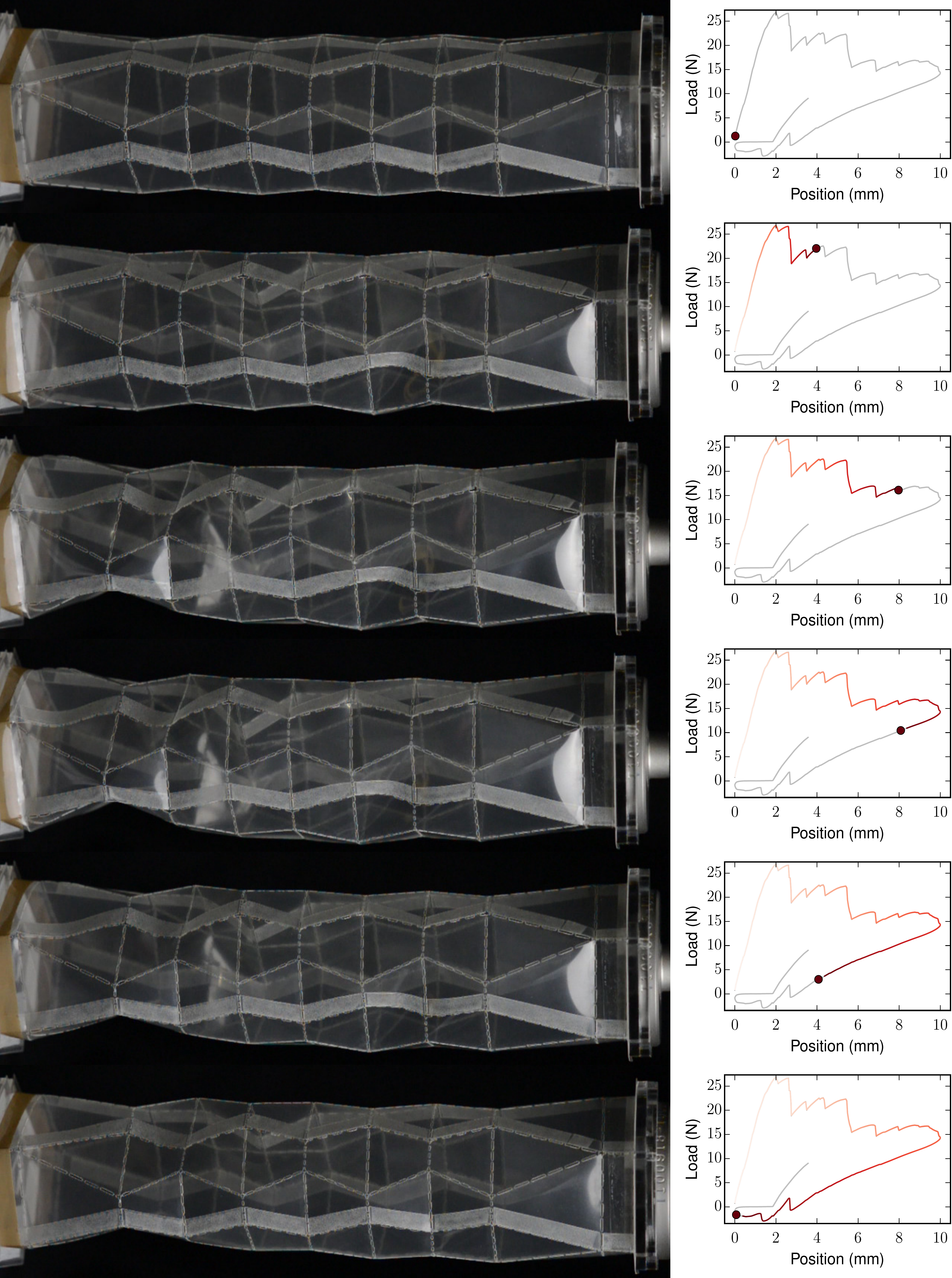}
\caption{Class~1--Type~B bellows with \(\phi_1=\ang{105}\) and \(\phi_2=\ang{69}\). This hexagonal bellows is expected to have a single accessible rigid-face configuration. Without a geometrically allowed collapsed state, this bellows buckles under load, breaking its rotational symmetry. See also Movie-S5 in~\cite{Note1}.}
\label{fig:bellows:hex:crumple}
\end{figure}

\begin{figure}[!htbp]
\includegraphics[width=0.5\textwidth]{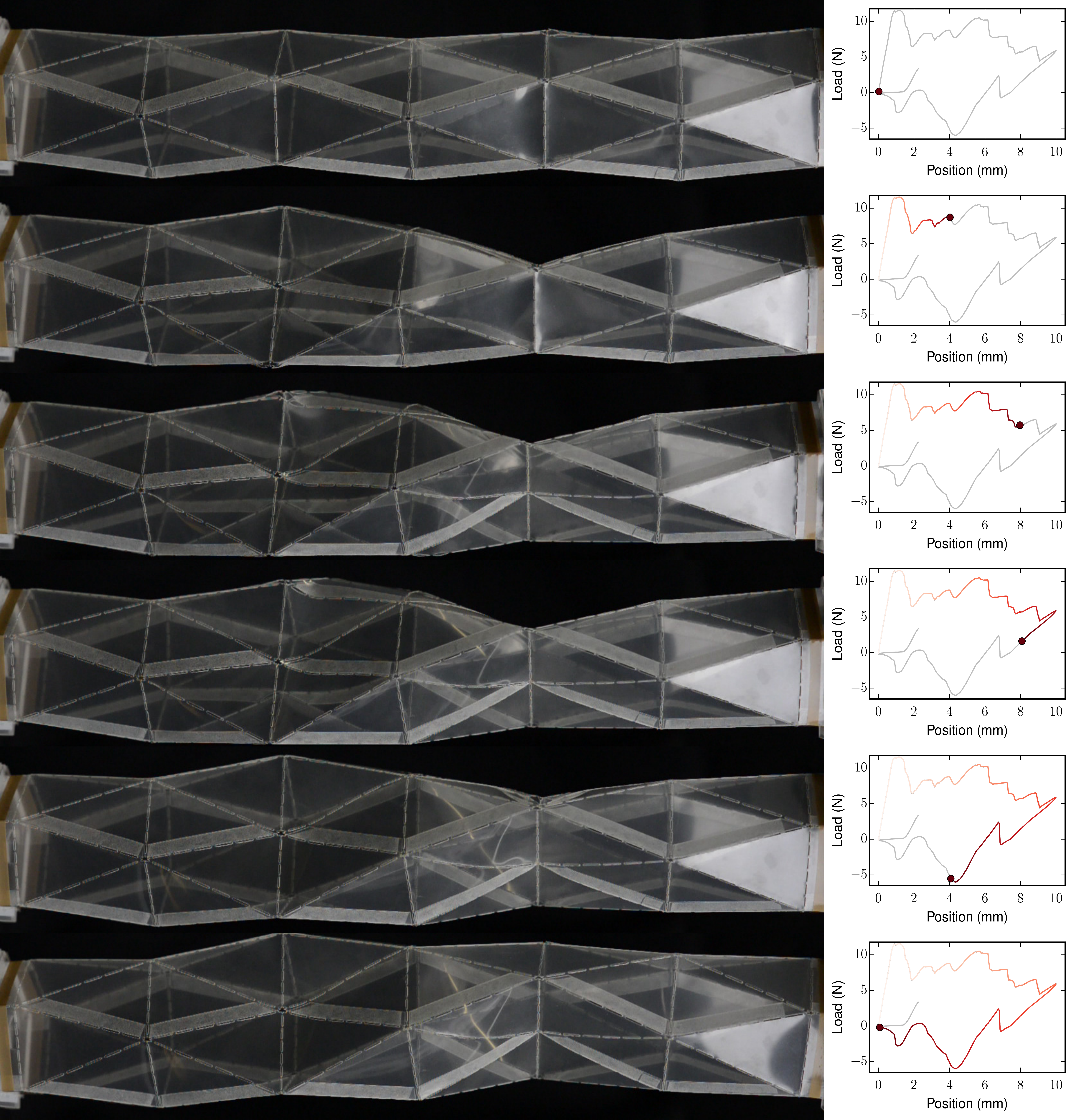}
\caption{Class~2-Type~B bellows with \(\phi_1=\ang{105}\) and \(\phi_2=\ang{69}\). Much like the hexagonal bellows of \cref{fig:bellows:hex:crumple}, without a geometrically allowed collapsed state this bellows crumples, breaking its rotational symmetry. See also Movie-S6 in~\cite{Note1}.}
\label{fig:bellows:yoshi:crumple}
\end{figure}

The mechanical response of the bellows was performed using an Instron$^\text{\textregistered}$ test frame and its corresponding deformation was captured with a Nikon D800 both as high resolution still images and as movie shots at 720p, 60 fps. Movies (available as supplemental material~\footnote{See Supplemental Material at http://link.aps.org/supplemental/ for movies of the mechanical testings corresponding to \cref{fig:bellows:phi68:miura} (Movie-S1), \cref{fig:bellows:phi68:tri} (Movie-S2), \cref{fig:bellows:phi76:miura} (Movie-S3), \cref{fig:bellows:phi76:tri} (Movie-S4), \cref{fig:bellows:hex:crumple} (Movie-S5) and \cref{fig:bellows:yoshi:crumple} (Movie-S6)}) were synchronized to the Instron$^\text{\textregistered}$ test frame's data after their capture via motion tracking. On all the following force versus displacement curves, the zero displacement point is chosen at the extended position of the bellows where the applied force is nearest to zero after attaching the bellows. The zero load reference point is established by the dynamic force value at the moment the test frame reverses direction from compression to extension. Notice that the first compression tends to have more dramatic behavior than the subsequent cycles. This atypical behavior is not reported and the figures below only illustrate the later compression and extension cycles. After the initial compression, subsequent cycles follow each other closely, with only minor aging of the device as it cycles.

The class~1--type~A bellows with \(\phi_1=\ang{68} <\phi_c\) in \cref{fig:bellows:phi68:miura} is the softest of the bellows examined here. As seen in its force-displacment curve as well as Movie-S1 in~\cite{Note1}, it can be fully collapsed with only \SI{5}{N} applied force. Unfortunately, we are unable to resolve face-bending behavior during the collapse of this pattern. Its class~2--type~A equivalent (\cref{fig:bellows:phi68:tri} and Movie-S2 in~\cite{Note1}) does show small steps as the paired cells collapse, but the smooth extension stroke (the lower half of the force-displacement plot) indicates that this pattern does not lock in its collapsed state, instead deploying itself after collapse.

The class~1--type~A bellows with \(\phi_1=\ang{76}>\phi_c\) in \cref{fig:bellows:phi76:miura} exhibits minor locking (see also Movie-S3 in~\cite{Note1}), but this is an artifact of the transition end pieces, which convert the variable concave cross-section to a uniform convex hexagon that can be mounted rigidly to the Instron$^\text{\textregistered}$. The class~2--Type~A bellows in \cref{fig:bellows:phi76:tri} with \(\phi_1=\ang{76}>\phi_c\) demonstrates the most interesting behavior of this set: clear face bending, crease bending, and self-locking behavior. An overview of its cycling behavior is displayed in Movie-S4~\cite{Note1}. The sharp steps on the extension stroke indicate locking of individual cells and the failure of those cells as they extend to their other rigid-faced state.

Looking closely at the ends of the bellows in the last frame of \cref{fig:bellows:phi76:tri} reveals that the outer edges parallel to \(\vv{\omega_2}\) (see \cref{tab:mother:alltypes}) are bent. The bellows has been driven out of its rigid face equilibrium, and the only way to access the stable collapsed state is by bending the faces. Class~2 bellows, however, are triangular and can only bend their faces by also bending a crease. The regularity of this crease bending is evidence of the suitability of the construction techniques detailed earlier. Crease bending also is far more energetically expensive than plate bending, which largely accounts for why the class~2 bellows are so much more rigid than their class~1 counterparts. After the crease bend emerges, it does not remain in the same location. By traversing the face, the vertex of the crease bend introduces a mobile crease to the fold pattern, thereby bypassing the preconditions required for the bellows theorem to apply~\cite{Guest:1994ty}.

In all cases, it is easier to extend a collapsed bellows than it is to collapse a rigidly extended bellows. This is readily explained by the interplay between the hinge bending and plate bending: When extended, the hinge is held relatively near its preferred angle of repose. The plate's dramatically higher folding energy then dominates the dynamics of the device, leading it to equilibrate near its rigid face state. On the hand, when collapsed, the hinges are compressed far from their angle of repose, leading to dramatic deformations of the plates, as seen in \cref{fig:bellows:phi76:tri}. Bent, the plates are driven away from their rigid-face equilibrium towards the potential barrier between the two states. Because the collapsed state is driven far from its equilibrium state by hinge energy, it is easier to overcome the remainder of the barrier.

Special attention should be paid to the mechanical response shown in~\cref{fig:bellows:phi76:tri}. In particular, note that there are three collapse events, with each one corresponding to a pair of panels collapsing. If the unit cells were identically manufactured and unable to communicate with one another, these progressive collapsing events would occur at roughly the same force levels. That they do not is potentially evidence that some cells are less rigid than others, due to minor variations in their manufacture. Moreover, careful inspection of video data Movie-S4 in~\cite{Note1} shows that the cells do not always collapse in the same order, suggesting that while some cells may be temporarily softer than others, it is not entirely explainable by their permanent connections to one another. Instead, we suspect that the remaining deployed cells are able to flex and deform slightly, easing the transition for the cells that will collapse. Because the number of available face configurations falls off as cells collapse, we see that each subsequent event requires more and more force.

Finally, bellows without a geometrically allowed collapsed state for both classes behave like typical thin-walled cylinders. As seen in \cref{fig:bellows:yoshi:crumple,fig:bellows:hex:crumple}, once the panels' capacity to deform is exceeded, the rotational symmetry breaks and one side of the bellows buckles and crumples in on itself. Returning the bellows to its original position does not restore the symmetry present initially (see also Movie-S5 and Movie-S6 in~\cite{Note1}).

Within these devices, the underlying mathematical deployability not only allows physical actuation, but is also largely enriched by the mechanics which in turn provides extended functional capabilities. Deployable designs with $\phi_1<\phi_c$ can be smoothly actuated with hardly any snap-through effects and with small forces: they can be held collapsed and they would self-deploy as the confining force is released. On the other hand, designs with $\phi_1>\phi_c$ will remain in a given metastable state until actuated to change configuration, and the available configurations can be made to be either fully collapsed versus fully deployed, or can alternatively exhibit several intermediate states with well defined energy barriers between states. Interestingly, the nature of the energy barrier between metastable states depends on the geometry of the faces: quadrangular faces decouple the bending of the face and that of the crease, leading to soft actuation, whereas triangular face bending requires the adjacent creases to deform, yielding an overall much stiffer actuation.

\section{Discussion}
\label{sec:discussion}

We have presented a unified geometrical description of several previously disconnected classes of origami bellows, and uncovered their allowed states, painting a phase diagram with a peculiar island of bistability. We then proceeded to craft the corresponding physical models using an original technique, which allowed us to investigate the effect of these kinematics on the mechanical response of the models. We have shown that it is possible to rapidly generate precisely folded origami bellows using a laser cutter, plastic film, and double-sided tape. This technique can be easily used to make arbitrarily complicated bellows with finely tuned fold parameters. We found that the ``deployability'' of a given pattern, a characterization of the distance between two geometrically allowed states, yields quantitative insight into the behavior of the physical bellows, though the mechanics of face and crease bending and self-exclusion govern the detailed characteristics of their deployment.    

More importantly, we have demonstrated the existence of a critical design angle that controls the bistability of a mechanical origami bellows. Furthermore, nonzero hinge energy drives these bistable devices away from the flat-folded configuration predicted by geometry, clearly illustrating the necessity of mechanics in a design phase. Class~1 bellows tend to be much more flexible than their Class~2 counterparts. As would be expected, the figures with larger \(\Delta\psi\) have more stable collapsed configurations, and some are capable of holding themselves in a collapsed state.

Bellows with specific properties can be designed with knowledge of the geometric limitations as well as the mechanical properties of their substrate. While the fabrication technique described earlier is optimal for exploring bellows configurations, it necessarily generates perforated bellows which are unable to displace fluids or be actuated by internal pressure. Construction of a sealed bellows is more complicated, requiring machined intermediate forms and molds to shape the cylindrical substrate as needed. It is thus a more efficient use of time to only generate these construction tools once a design has been adequately prototyped by using our paper lantern construction technique.

Given the importance of plate-bending in the collapse of an origami-folded bellows and the difficulty faced by finite-element computations of thin shells undergoing bending~\cite{Chapelle:1998ux}, physical measurements of origami patterned bellows as they collapse or deploy are crucial for the verification and development of simulation tools. As faces and hinges bend, accurate measurements of surface curvature would allow a more complete accounting of the effective mechanical response of the bellows.

Among promising fields of application for next generation bellows design are architecture, mechanical design, and cryogenic devices. In these arenas, reliability and weight are primary design goals in the development of deployable origami bellows. Materials age and fail dramatically faster when subjected to crease bending, so patterns that minimize this damage are more reliable. With a full understanding of the importance of face bending, more robust origami patterned actuators may be designed. Bellows designs to date have rightfully started with purely geometric considerations, followed by iterative physical prototyping. Unfortunately, as the practical foldability of a bellows depends on the interplay between the established creases and the device's ability to plate-bend its faces, this process is dependent on prototyping and intuition to generate future designs with predetermined functionalities.

\begin{acknowledgments}
This material is based upon research supported by the Chateaubriand Fellowship of the Office for Science \& Technology of the Embassy of France in the United States (AR), french national research agency grant ANR-14-CE07-0031 METAMAT (FL) and FONDECYT (Chile) grant No.~1130709 (SR).
\end{acknowledgments}

\bibliography{Article.bib}

\begin{thebibliography}{28}%
\makeatletter
\providecommand \@ifxundefined [1]{%
 \@ifx{#1\undefined}
}%
\providecommand \@ifnum [1]{%
 \ifnum #1\expandafter \@firstoftwo
 \else \expandafter \@secondoftwo
 \fi
}%
\providecommand \@ifx [1]{%
 \ifx #1\expandafter \@firstoftwo
 \else \expandafter \@secondoftwo
 \fi
}%
\providecommand \natexlab [1]{#1}%
\providecommand \enquote  [1]{``#1''}%
\providecommand \bibnamefont  [1]{#1}%
\providecommand \bibfnamefont [1]{#1}%
\providecommand \citenamefont [1]{#1}%
\providecommand \href@noop [0]{\@secondoftwo}%
\providecommand \href [0]{\begingroup \@sanitize@url \@href}%
\providecommand \@href[1]{\@@startlink{#1}\@@href}%
\providecommand \@@href[1]{\endgroup#1\@@endlink}%
\providecommand \@sanitize@url [0]{\catcode `\\12\catcode `\$12\catcode
  `\&12\catcode `\#12\catcode `\^12\catcode `\_12\catcode `\%12\relax}%
\providecommand \@@startlink[1]{}%
\providecommand \@@endlink[0]{}%
\providecommand \url  [0]{\begingroup\@sanitize@url \@url }%
\providecommand \@url [1]{\endgroup\@href {#1}{\urlprefix }}%
\providecommand \urlprefix  [0]{URL }%
\providecommand \Eprint [0]{\href }%
\providecommand \doibase [0]{http://dx.doi.org/}%
\providecommand \selectlanguage [0]{\@gobble}%
\providecommand \bibinfo  [0]{\@secondoftwo}%
\providecommand \bibfield  [0]{\@secondoftwo}%
\providecommand \translation [1]{[#1]}%
\providecommand \BibitemOpen [0]{}%
\providecommand \bibitemStop [0]{}%
\providecommand \bibitemNoStop [0]{.\EOS\space}%
\providecommand \EOS [0]{\spacefactor3000\relax}%
\providecommand \BibitemShut  [1]{\csname bibitem#1\endcsname}%
\let\auto@bib@innerbib\@empty
\bibitem [{\citenamefont {Leb{\'e}e}(2015)}]{lebee2015folds}%
  \BibitemOpen
  \bibfield  {author} {\bibinfo {author} {\bibfnamefont {A.}~\bibnamefont
  {Leb{\'e}e}},\ }\href@noop {} {\bibfield  {journal} {\bibinfo  {journal}
  {International Journal of Space Structures}\ }\textbf {\bibinfo {volume}
  {30}},\ \bibinfo {pages} {55} (\bibinfo {year} {2015})}\BibitemShut {NoStop}%
\bibitem [{\citenamefont {Schenk}\ and\ \citenamefont
  {Guest}(2013)}]{schenk2013geometry}%
  \BibitemOpen
  \bibfield  {author} {\bibinfo {author} {\bibfnamefont {M.}~\bibnamefont
  {Schenk}}\ and\ \bibinfo {author} {\bibfnamefont {S.~D.}\ \bibnamefont
  {Guest}},\ }\href@noop {} {\bibfield  {journal} {\bibinfo  {journal}
  {Proceedings of the National Academy of Sciences}\ }\textbf {\bibinfo
  {volume} {110}},\ \bibinfo {pages} {3276} (\bibinfo {year}
  {2013})}\BibitemShut {NoStop}%
\bibitem [{\citenamefont {Silverberg}\ \emph {et~al.}(2014)\citenamefont
  {Silverberg}, \citenamefont {Evans}, \citenamefont {McLeod}, \citenamefont
  {Hayward}, \citenamefont {Hull}, \citenamefont {Santangelo},\ and\
  \citenamefont {Cohen}}]{silverberg2014using}%
  \BibitemOpen
  \bibfield  {author} {\bibinfo {author} {\bibfnamefont {J.~L.}\ \bibnamefont
  {Silverberg}}, \bibinfo {author} {\bibfnamefont {A.~A.}\ \bibnamefont
  {Evans}}, \bibinfo {author} {\bibfnamefont {L.}~\bibnamefont {McLeod}},
  \bibinfo {author} {\bibfnamefont {R.~C.}\ \bibnamefont {Hayward}}, \bibinfo
  {author} {\bibfnamefont {T.}~\bibnamefont {Hull}}, \bibinfo {author}
  {\bibfnamefont {C.~D.}\ \bibnamefont {Santangelo}}, \ and\ \bibinfo {author}
  {\bibfnamefont {I.}~\bibnamefont {Cohen}},\ }\href@noop {} {\bibfield
  {journal} {\bibinfo  {journal} {Science}\ }\textbf {\bibinfo {volume}
  {345}},\ \bibinfo {pages} {647} (\bibinfo {year} {2014})}\BibitemShut
  {NoStop}%
\bibitem [{\citenamefont {Wei}\ \emph {et~al.}(2013)\citenamefont {Wei},
  \citenamefont {Guo}, \citenamefont {Dudte}, \citenamefont {Liang},\ and\
  \citenamefont {Mahadevan}}]{Wei:2013kn}%
  \BibitemOpen
  \bibfield  {author} {\bibinfo {author} {\bibfnamefont {Z.~Y.}\ \bibnamefont
  {Wei}}, \bibinfo {author} {\bibfnamefont {Z.~V.}\ \bibnamefont {Guo}},
  \bibinfo {author} {\bibfnamefont {L.}~\bibnamefont {Dudte}}, \bibinfo
  {author} {\bibfnamefont {H.~Y.}\ \bibnamefont {Liang}}, \ and\ \bibinfo
  {author} {\bibfnamefont {L.}~\bibnamefont {Mahadevan}},\ }\href@noop {}
  {\bibfield  {journal} {\bibinfo  {journal} {Physical Review Letters}\
  }\textbf {\bibinfo {volume} {110}},\ \bibinfo {pages} {215501} (\bibinfo
  {year} {2013})}\BibitemShut {NoStop}%
\bibitem [{\citenamefont {Silverberg}\ \emph {et~al.}(2015)\citenamefont
  {Silverberg}, \citenamefont {Na}, \citenamefont {Evans}, \citenamefont {Liu},
  \citenamefont {Hull}, \citenamefont {Santangelo}, \citenamefont {Lang},
  \citenamefont {Hayward},\ and\ \citenamefont {Cohen}}]{Silverberg:2015gb}%
  \BibitemOpen
  \bibfield  {author} {\bibinfo {author} {\bibfnamefont {J.~L.}\ \bibnamefont
  {Silverberg}}, \bibinfo {author} {\bibfnamefont {J.-H.}\ \bibnamefont {Na}},
  \bibinfo {author} {\bibfnamefont {A.~A.}\ \bibnamefont {Evans}}, \bibinfo
  {author} {\bibfnamefont {B.}~\bibnamefont {Liu}}, \bibinfo {author}
  {\bibfnamefont {T.~C.}\ \bibnamefont {Hull}}, \bibinfo {author}
  {\bibfnamefont {C.~D.}\ \bibnamefont {Santangelo}}, \bibinfo {author}
  {\bibfnamefont {R.~J.}\ \bibnamefont {Lang}}, \bibinfo {author}
  {\bibfnamefont {R.~C.}\ \bibnamefont {Hayward}}, \ and\ \bibinfo {author}
  {\bibfnamefont {I.}~\bibnamefont {Cohen}},\ }\href@noop {} {\bibfield
  {journal} {\bibinfo  {journal} {Nature Materials}\ }\textbf {\bibinfo
  {volume} {14}},\ \bibinfo {pages} {389} (\bibinfo {year} {2015})}\BibitemShut
  {NoStop}%
\bibitem [{\citenamefont {Brunck}\ \emph {et~al.}(2016)\citenamefont {Brunck},
  \citenamefont {Lechenault}, \citenamefont {Reid},\ and\ \citenamefont
  {Adda-Bedia}}]{Brunck:2016jr}%
  \BibitemOpen
  \bibfield  {author} {\bibinfo {author} {\bibfnamefont {V.}~\bibnamefont
  {Brunck}}, \bibinfo {author} {\bibfnamefont {F.}~\bibnamefont {Lechenault}},
  \bibinfo {author} {\bibfnamefont {A.}~\bibnamefont {Reid}}, \ and\ \bibinfo
  {author} {\bibfnamefont {M.}~\bibnamefont {Adda-Bedia}},\ }\href@noop {}
  {\bibfield  {journal} {\bibinfo  {journal} {Physical Review E}\ }\textbf
  {\bibinfo {volume} {93}},\ \bibinfo {pages} {033005} (\bibinfo {year}
  {2016})}\BibitemShut {NoStop}%
\bibitem [{\citenamefont {Demaine}(2001)}]{Demaine:2001uq}%
  \BibitemOpen
  \bibfield  {author} {\bibinfo {author} {\bibfnamefont {E.~D.}\ \bibnamefont
  {Demaine}},\ }\emph {\bibinfo {title} {{Folding and unfolding}}},\ \href@noop
  {} {Ph.D. thesis},\ \bibinfo  {school} {University of Waterloo} (\bibinfo
  {year} {2001})\BibitemShut {NoStop}%
\bibitem [{\citenamefont {Belcastro}\ and\ \citenamefont
  {Hull}(2002)}]{Belcastro:2002ku}%
  \BibitemOpen
  \bibfield  {author} {\bibinfo {author} {\bibfnamefont {S.~M.}\ \bibnamefont
  {Belcastro}}\ and\ \bibinfo {author} {\bibfnamefont {T.~C.}\ \bibnamefont
  {Hull}},\ }\href@noop {} {\bibfield  {journal} {\bibinfo  {journal} {Linear
  Algebra and Its Applications}\ }\textbf {\bibinfo {volume} {348}},\ \bibinfo
  {pages} {273} (\bibinfo {year} {2002})}\BibitemShut {NoStop}%
\bibitem [{\citenamefont {Dureisseix}(2012)}]{Dureisseix:2012tk}%
  \BibitemOpen
  \bibfield  {author} {\bibinfo {author} {\bibfnamefont {D.}~\bibnamefont
  {Dureisseix}},\ }\href@noop {} {\bibfield  {journal} {\bibinfo  {journal}
  {International Journal of Space Structures}\ }\textbf {\bibinfo {volume}
  {27}},\ \bibinfo {pages} {1} (\bibinfo {year} {2012})}\BibitemShut {NoStop}%
\bibitem [{\citenamefont {Lechenault}\ \emph {et~al.}(2014)\citenamefont
  {Lechenault}, \citenamefont {Thiria},\ and\ \citenamefont
  {Adda-Bedia}}]{Lechenault:2014wn}%
  \BibitemOpen
  \bibfield  {author} {\bibinfo {author} {\bibfnamefont {F.}~\bibnamefont
  {Lechenault}}, \bibinfo {author} {\bibfnamefont {B.}~\bibnamefont {Thiria}},
  \ and\ \bibinfo {author} {\bibfnamefont {M.}~\bibnamefont {Adda-Bedia}},\
  }\href@noop {} {\bibfield  {journal} {\bibinfo  {journal} {Physical Review
  Letters}\ }\textbf {\bibinfo {volume} {112}},\ \bibinfo {pages} {244301}
  (\bibinfo {year} {2014})}\BibitemShut {NoStop}%
\bibitem [{\citenamefont {Schenk}\ \emph {et~al.}(2014)\citenamefont {Schenk},
  \citenamefont {Viquerat}, \citenamefont {Seffen},\ and\ \citenamefont
  {Guest}}]{Schenk:2014jc}%
  \BibitemOpen
  \bibfield  {author} {\bibinfo {author} {\bibfnamefont {M.}~\bibnamefont
  {Schenk}}, \bibinfo {author} {\bibfnamefont {A.~D.}\ \bibnamefont
  {Viquerat}}, \bibinfo {author} {\bibfnamefont {K.~A.}\ \bibnamefont
  {Seffen}}, \ and\ \bibinfo {author} {\bibfnamefont {S.~D.}\ \bibnamefont
  {Guest}},\ }\href@noop {} {\bibfield  {journal} {\bibinfo  {journal} {Journal
  of Spacecraft and Rockets}\ }\textbf {\bibinfo {volume} {51}},\ \bibinfo
  {pages} {762} (\bibinfo {year} {2014})}\BibitemShut {NoStop}%
\bibitem [{\citenamefont {Guest}(1994)}]{Guest:1994uo}%
  \BibitemOpen
  \bibfield  {author} {\bibinfo {author} {\bibfnamefont {S.~D.}\ \bibnamefont
  {Guest}},\ }\emph {\bibinfo {title} {{Deployable structures: concepts and
  analysis}}},\ \href@noop {} {Ph.D. thesis},\ \bibinfo  {school} {University
  of Cambridge} (\bibinfo {year} {1994})\BibitemShut {NoStop}%
\bibitem [{\citenamefont {Kuribayashi}\ \emph {et~al.}(2006)\citenamefont
  {Kuribayashi}, \citenamefont {Tsuchiya}, \citenamefont {You}, \citenamefont
  {Tomus}, \citenamefont {Umemoto}, \citenamefont {Ito},\ and\ \citenamefont
  {Sasaki}}]{Kuribayashi:2006wo}%
  \BibitemOpen
  \bibfield  {author} {\bibinfo {author} {\bibfnamefont {K.}~\bibnamefont
  {Kuribayashi}}, \bibinfo {author} {\bibfnamefont {K.}~\bibnamefont
  {Tsuchiya}}, \bibinfo {author} {\bibfnamefont {Z.}~\bibnamefont {You}},
  \bibinfo {author} {\bibfnamefont {D.}~\bibnamefont {Tomus}}, \bibinfo
  {author} {\bibfnamefont {M.}~\bibnamefont {Umemoto}}, \bibinfo {author}
  {\bibfnamefont {T.}~\bibnamefont {Ito}}, \ and\ \bibinfo {author}
  {\bibfnamefont {M.}~\bibnamefont {Sasaki}},\ }\href@noop {} {\bibfield
  {journal} {\bibinfo  {journal} {Materials Science and Engineering: A}\
  }\textbf {\bibinfo {volume} {419}},\ \bibinfo {pages} {131} (\bibinfo {year}
  {2006})}\BibitemShut {NoStop}%
\bibitem [{\citenamefont {Reid}()}]{Austin}%
  \BibitemOpen
  \bibfield  {author} {\bibinfo {author} {\bibfnamefont {A.}~\bibnamefont
  {Reid}},\ }\href@noop {} {}\bibinfo {note} {In preparation}\BibitemShut
  {NoStop}%
\bibitem [{\citenamefont {Schenk}\ \emph {et~al.}(2013)\citenamefont {Schenk},
  \citenamefont {Kerr}, \citenamefont {Smyth},\ and\ \citenamefont
  {Guest}}]{Schenk:2013to}%
  \BibitemOpen
  \bibfield  {author} {\bibinfo {author} {\bibfnamefont {M.}~\bibnamefont
  {Schenk}}, \bibinfo {author} {\bibfnamefont {S.}~\bibnamefont {Kerr}},
  \bibinfo {author} {\bibfnamefont {A.}~\bibnamefont {Smyth}}, \ and\ \bibinfo
  {author} {\bibfnamefont {S.}~\bibnamefont {Guest}},\ }in\ \href@noop {}
  {\emph {\bibinfo {booktitle} {Proceedings of the First Conference
  Transformables, Sept}}}\ (\bibinfo {year} {2013})\ pp.\ \bibinfo {pages}
  {18--20}\BibitemShut {NoStop}%
\bibitem [{\citenamefont {Tachi}\ and\ \citenamefont
  {Miura}(2012)}]{Tachi:2012wl}%
  \BibitemOpen
  \bibfield  {author} {\bibinfo {author} {\bibfnamefont {T.}~\bibnamefont
  {Tachi}}\ and\ \bibinfo {author} {\bibfnamefont {K.}~\bibnamefont {Miura}},\
  }\href@noop {} {\bibfield  {journal} {\bibinfo  {journal} {J. Int. Assoc.
  Shell Spat. Struct}\ }\textbf {\bibinfo {volume} {53}},\ \bibinfo {pages}
  {217} (\bibinfo {year} {2012})}\BibitemShut {NoStop}%
\bibitem [{\citenamefont {Yasuda}\ and\ \citenamefont
  {Yang}(2015)}]{Yasuda:2015vy}%
  \BibitemOpen
  \bibfield  {author} {\bibinfo {author} {\bibfnamefont {H.}~\bibnamefont
  {Yasuda}}\ and\ \bibinfo {author} {\bibfnamefont {J.}~\bibnamefont {Yang}},\
  }\href@noop {} {\bibfield  {journal} {\bibinfo  {journal} {Physical Review
  Letters}\ }\textbf {\bibinfo {volume} {114}},\ \bibinfo {pages} {185502}
  (\bibinfo {year} {2015})}\BibitemShut {NoStop}%
\bibitem [{\citenamefont {B{\"o}s}\ \emph {et~al.}(2015)\citenamefont
  {B{\"o}s}, \citenamefont {Vouga}, \citenamefont {Gottesman},\ and\
  \citenamefont {Wardetzky}}]{Bos:2015uu}%
  \BibitemOpen
  \bibfield  {author} {\bibinfo {author} {\bibfnamefont {F.}~\bibnamefont
  {B{\"o}s}}, \bibinfo {author} {\bibfnamefont {E.}~\bibnamefont {Vouga}},
  \bibinfo {author} {\bibfnamefont {O.}~\bibnamefont {Gottesman}}, \ and\
  \bibinfo {author} {\bibfnamefont {M.}~\bibnamefont {Wardetzky}},\ }\href@noop
  {} {\bibfield  {journal} {\bibinfo  {journal} {arXiv.org}\ } (\bibinfo {year}
  {2015})},\ \Eprint {http://arxiv.org/abs/1507.08472v1} {1507.08472v1}
  \BibitemShut {NoStop}%
\bibitem [{\citenamefont {Cai}\ \emph {et~al.}(2015)\citenamefont {Cai},
  \citenamefont {Deng}, \citenamefont {Feng},\ and\ \citenamefont
  {Zhou}}]{Cai:2015hb}%
  \BibitemOpen
  \bibfield  {author} {\bibinfo {author} {\bibfnamefont {J.}~\bibnamefont
  {Cai}}, \bibinfo {author} {\bibfnamefont {X.}~\bibnamefont {Deng}}, \bibinfo
  {author} {\bibfnamefont {J.}~\bibnamefont {Feng}}, \ and\ \bibinfo {author}
  {\bibfnamefont {Y.}~\bibnamefont {Zhou}},\ }\href@noop {} {\bibfield
  {journal} {\bibinfo  {journal} {Smart Materials and Structures}\ }\textbf
  {\bibinfo {volume} {24}},\ \bibinfo {pages} {1} (\bibinfo {year}
  {2015})}\BibitemShut {NoStop}%
\bibitem [{\citenamefont {Jianguo}\ \emph {et~al.}(2015)\citenamefont
  {Jianguo}, \citenamefont {Xiaowei}, \citenamefont {Ya}, \citenamefont
  {Jian},\ and\ \citenamefont {Yongming}}]{Jianguo:2015ic}%
  \BibitemOpen
  \bibfield  {author} {\bibinfo {author} {\bibfnamefont {C.}~\bibnamefont
  {Jianguo}}, \bibinfo {author} {\bibfnamefont {D.}~\bibnamefont {Xiaowei}},
  \bibinfo {author} {\bibfnamefont {Z.}~\bibnamefont {Ya}}, \bibinfo {author}
  {\bibfnamefont {F.}~\bibnamefont {Jian}}, \ and\ \bibinfo {author}
  {\bibfnamefont {T.}~\bibnamefont {Yongming}},\ }\href@noop {} {\bibfield
  {journal} {\bibinfo  {journal} {Journal of Mechanical Design}\ }\textbf
  {\bibinfo {volume} {137}},\ \bibinfo {pages} {061406} (\bibinfo {year}
  {2015})}\BibitemShut {NoStop}%
\bibitem [{\citenamefont {Wu}\ and\ \citenamefont {You}(2010)}]{Wu:2010gq}%
  \BibitemOpen
  \bibfield  {author} {\bibinfo {author} {\bibfnamefont {W.}~\bibnamefont
  {Wu}}\ and\ \bibinfo {author} {\bibfnamefont {Z.}~\bibnamefont {You}},\
  }\href@noop {} {\bibfield  {journal} {\bibinfo  {journal} {Proceedings of the
  Royal Society A: Mathematical, Physical and Engineering Sciences}\ }\textbf
  {\bibinfo {volume} {466}},\ \bibinfo {pages} {2155} (\bibinfo {year}
  {2010})}\BibitemShut {NoStop}%
\bibitem [{\citenamefont {Yoshimura}(1955)}]{Yoshimura:1955up}%
  \BibitemOpen
  \bibfield  {author} {\bibinfo {author} {\bibfnamefont {Y.}~\bibnamefont
  {Yoshimura}},\ }\href@noop {} {\emph {\bibinfo {title} {{On the mechanism of
  buckling of a circular cylindrical shell under axial compression}}}},\
  \bibinfo {type} {Tech. Rep.}\ \bibinfo {number} {1390}\ (\bibinfo
  {institution} {National Advisory Committee for Aeronautics},\ \bibinfo {year}
  {1955})\BibitemShut {NoStop}%
\bibitem [{\citenamefont {Hull}(2002)}]{Hull:2002wn}%
  \BibitemOpen
  \bibfield  {author} {\bibinfo {author} {\bibfnamefont {T.~C.}\ \bibnamefont
  {Hull}},\ }\enquote {\bibinfo {title} {{Origami 3: Proceedings of the Third
  International Meeting of Origami Science}},}\ \ (\bibinfo  {publisher}
  {Mathematics},\ \bibinfo {year} {2002})\ pp.\ \bibinfo {pages}
  {197--207}\BibitemShut {NoStop}%
\bibitem [{\citenamefont {Connelly}\ \emph {et~al.}(1997)\citenamefont
  {Connelly}, \citenamefont {Sabitov},\ and\ \citenamefont
  {Walz}}]{Connelly:1997tw}%
  \BibitemOpen
  \bibfield  {author} {\bibinfo {author} {\bibfnamefont {R.}~\bibnamefont
  {Connelly}}, \bibinfo {author} {\bibfnamefont {I.}~\bibnamefont {Sabitov}}, \
  and\ \bibinfo {author} {\bibfnamefont {A.}~\bibnamefont {Walz}},\ }\href@noop
  {} {\bibfield  {journal} {\bibinfo  {journal} {Beitr\"{a}ge zur Algebra und
  Geometrie}\ }\textbf {\bibinfo {volume} {38}},\ \bibinfo {pages} {1}
  (\bibinfo {year} {1997})}\BibitemShut {NoStop}%
\bibitem [{\citenamefont {Kane}(2000)}]{Kane:2000tf}%
  \BibitemOpen
  \bibfield  {author} {\bibinfo {author} {\bibfnamefont {N.~R.}\ \bibnamefont
  {Kane}},\ }\href@noop {} {\enquote {\bibinfo {title} {{Mathematically
  optimized family of ultra low distortion bellow fold patterns}},}\ }\bibinfo
  {howpublished} {US Patent 6,054,194} (\bibinfo {year} {2000})\BibitemShut
  {NoStop}%
\bibitem [{Note1()}]{Note1}%
  \BibitemOpen
  \bibinfo {note} {See Supplemental Material at
  http://link.aps.org/supplemental/ for movies of the mechanical testings
  corresponding to \protect \cref {fig:bellows:phi68:miura} (Movie-S1),
  \protect \cref {fig:bellows:phi68:tri} (Movie-S2), \protect \cref
  {fig:bellows:phi76:miura} (Movie-S3), \protect \cref {fig:bellows:phi76:tri}
  (Movie-S4), \protect \cref {fig:bellows:hex:crumple} (Movie-S5) and \protect
  \cref {fig:bellows:yoshi:crumple} (Movie-S6)}\BibitemShut {NoStop}%
\bibitem [{\citenamefont {Guest}\ and\ \citenamefont
  {Pellegrino}(1994)}]{Guest:1994ty}%
  \BibitemOpen
  \bibfield  {author} {\bibinfo {author} {\bibfnamefont {S.~D.}\ \bibnamefont
  {Guest}}\ and\ \bibinfo {author} {\bibfnamefont {S.}~\bibnamefont
  {Pellegrino}},\ }\href@noop {} {\bibfield  {journal} {\bibinfo  {journal}
  {Journal of Applied Mechanics}\ }\textbf {\bibinfo {volume} {61}},\ \bibinfo
  {pages} {773} (\bibinfo {year} {1994})}\BibitemShut {NoStop}%
\bibitem [{\citenamefont {Chapelle}\ and\ \citenamefont
  {Bathe}(1998)}]{Chapelle:1998ux}%
  \BibitemOpen
  \bibfield  {author} {\bibinfo {author} {\bibfnamefont {D.}~\bibnamefont
  {Chapelle}}\ and\ \bibinfo {author} {\bibfnamefont {K.-J.}\ \bibnamefont
  {Bathe}},\ }\href@noop {} {\bibfield  {journal} {\bibinfo  {journal}
  {Computers {\&} Structures}\ }\textbf {\bibinfo {volume} {66}},\ \bibinfo
  {pages} {19} (\bibinfo {year} {1998})}\BibitemShut {NoStop}%
\end{thebibliography}%

\end{document}